\documentclass{amsart}    
\usepackage{amssymb}    
\usepackage{latexsym}    
    
\newcommand{\ZZ}{{\mathbb Z}}    
\newcommand{\RR}{{\mathbb R}}    
\newcommand{\CC}{{\mathbb C}}    
\newcommand{\NN}{{\mathbb N}}

\newtheorem{theorem}{Theorem}       
\newtheorem{lemma}{Lemma}[section]       
\newtheorem{prop}[lemma]{Proposition}       
\newtheorem{coro}[lemma]{Corollary}       
\newtheorem{definition}[lemma]{Definition}       
\renewcommand{\Im}{{\rm Im}}    
    
\newcommand{\tr}{{\mathrm{tr}}}    
    
\newcounter{smalllist}    
\newenvironment{SmallList}{%
\begin{list}{\roman{smalllist})}%
{\setlength{\topsep}{0mm}\setlength{\parsep}{0mm}\setlength{\itemsep}{0mm}%
\setlength{\labelwidth}{10mm}\usecounter{smalllist}}%
}{\end{list}}    
    
\newenvironment{CaseHier}[1]{%
\begin{list}{}%
{\setlength{\topsep}{#1}\setlength{\parsep}{0mm}\setlength{\itemsep}{#1}%
\setlength{\labelwidth}{0mm}\setlength{\labelsep}{0mm}\setlength{\itemindent}{0mm}\setlength{\leftmargin}{0mm}}%
}{\end{list}}    
    
\begin{document}    
\title[Uniform {\larger[1]$\alpha$}-continuity for 1D quasicrystals]{Uniform spectral properties of one-dimensional quasicrystals, III.    
{\larger[1]$\boldsymbol\alpha$}-continuity}    
\author[D.~Damanik, R.~Killip, D.~Lenz]{David Damanik$\,^{1,2}$, Rowan Killip$\,^{1}$, Daniel Lenz$\,^{2}$}    
\maketitle    
\vspace{0.3cm}        
\noindent        
$^1$ Department of Mathematics 253--37, California Institute of Technology,        
Pasadena, CA 91125, U.S.A.\\[0.2cm]        
$^2$ Fachbereich Mathematik, Johann Wolfgang Goethe-Universit\"at,        
60054 Frankfurt, Germany\\[0.3cm]        
\begin{abstract}    
We study the spectral properties of discrete one-dimensional Schr\"odinger operators with Sturmian potentials.    
It is shown that the point spectrum is always empty. Moreover, for rotation numbers with bounded density, we establish    
purely $\alpha$-continuous spectrum, uniformly for all phases. The proofs rely on the unique decomposition property of    
Sturmian potentials, a mass-reproduction technique based upon a Gordon-type argument, and on the Jitomirskaya-Last    
extension of the Gilbert-Pearson theory of subordinacy.     
\end{abstract}    
    
\section{Introduction}    
    
In this article we continue the study of the spectral properties of discrete one-dimensional Schr\"odinger operators with Sturmian potentials which was started in \cite{dl1,dl2}. That is, we shall consider the operators    
\begin{equation}\label{hlt}    
[H_{\lambda,\theta,\beta}u](n)=u(n+1)+u(n-1)+\lambda v_{\theta,\beta}(n)u(n),    
\end{equation}    
acting in $\ell^2(\ZZ)$. Here      
$$    
v_{\theta,\beta}(n)=\chi_{[1-\theta,1)}\big(n\theta +\beta\text{ mod $1$}\big),    
$$    
with coupling constant $\lambda \in \RR \setminus \{0\}$, irrational rotation number $\theta \in (0,1)$, and phase    
$\beta \in [0,1)$. Most of our study will be concerned with the investigation of the solutions to the corresponding difference equation    
\begin{equation}\label{eve}    
(H_{\lambda,\theta,\beta} - E) u = 0 ,    
\end{equation}     
where the sequence $(u(n))_{n \in \ZZ}$ will always be normalized in the sense that    
\begin{equation}\label{norm}    
|u(0)|^2+|u(1)|^2=1.    
\end{equation}    
    
The family of operators $(H_{\lambda,\theta,\beta})$ is commonly agreed to model a one-dimensional quasicrystal. It provides a natural generalization of the Fibonacci family of operators which corresponds to rotation number    
$\theta=\theta_F = \frac{\sqrt{5} - 1}{2}$, the golden mean. This model was introduced independently by two groups in the early 80's \cite{kkt,oprss} and has been studied extensively since. The review articles \cite{d2,s7} recount the history of generalizations of the basic Fibonacci model and    
the results obtained for each of them.    
    
It was conjectured early that the operators $H_{\lambda,\theta,\beta}$ have purely singular continuous zero-measure spectrum. In the Fibonacci case this belief was made explicit, for example, in \cite{ko}. A major step towards establishing these properties was achieved when it was realized that the general Kotani theory \cite{k2} has very strong consequences for potentials taking finitely many values \cite{k3}. This immediately implied (by combining it with a recent stability result of Last and Simon \cite{ls}) that the absolutely continuous spectrum of $H_{\lambda,\theta,\beta}$ is empty. Moreover, the key technical result of \cite{k3} was used by S\"ut\H{o} \cite{s6} and Bellissard et al.~\cite{bist} to establish the zero-measure property in full generality.    
That is, for all admissible parameter values, the spectrum of $H_{\lambda,\theta,\beta}$ has zero Lebesgue measure, and hence is a Cantor set. All previously known results on the absence of point spectrum, however, were partial and generic in either a topological or measure-theoretical sense. First, Delyon and Petritis proved empty point spectrum for every $\lambda$, almost every $\theta$, and almost every $\beta$ \cite{dp1}. S\"ut\H{o} then proved absence of eigenvalues in the Fibonacci case, which was not contained in the full measure set of rotation numbers from the Delyon-Petritis result, for $\beta = 0$ and all $\lambda$ \cite{s5}. By general principles, this yields empty point spectrum for a dense $G_\delta$-set of $\beta$'s \cite{s1}. This result was implicitly extended to arbitrary $\theta$ by Bellissard et al.~\cite{bist} (see \cite{d1,hks} for explicit proofs). In \cite{k1} Kaminaga proved  absence of eigenvalues for every $\lambda$, every $\theta$, and almost every $\beta$. Finally, the paper \cite{dl1} treated every $\lambda$, almost every $\theta$, and every $\beta$. Our first goal here is to complete the identification    
of the spectral type for all parameter values.    
    
\begin{theorem}\label{cont}    
For every $\lambda,\theta,\beta$, the operator $H_{\lambda,\theta,\beta}$ has empty point spectrum.    
\end{theorem}    
    
\begin{coro}    
For every $\lambda,\theta,\beta$, the operator $H_{\lambda,\theta,\beta}$ has purely singular continuous zero-measure spectrum.    
\end{coro}    
    
Recently, the understanding of quantum dynamics for operators with purely singular continuous spectrum has been    
considerably improved. Lower bounds on the time evolution of the associated quantum systems can be obtained by means of certain Hausdorff dimensional properties of the spectral measures. The first results in this direction were obtained by Guarneri \cite{g3} and Combes \cite{c2}. They essentially required uniform $\alpha$-H\"older continuity. Their results were extended by Last in \cite{l} to spectral measures with non-trivial $\alpha$-continuous component, that is, measures that are not supported on a set of zero $\alpha$-Hausdorff measure. We refer the reader to \cite{bcm,bt,gs} for subsequent developments. Apart from the implications for dynamics, the notion of $\alpha$-continuity has the advantage of being accessible by an investigation of solutions to (\ref{eve}) as was realized by Jitomirskaya and Last \cite{jl1,jl2,jl3}. In these papers they establish an extension of the classical Gilbert-Pearson theory of subordinacy \cite{g1,gp,kp} and present several applications. Using their general approach, \cite{d1} established purely $\alpha$-continuous spectral measures for the operators $H_{\lambda,\theta,\beta}$ for every $\lambda$, an uncountable zero-measure set of $\theta$'s and $\beta = 0$ with a positive $\alpha$ which depends on both $\lambda$ and $\theta$. From a physical point of view, however, the dynamical implications that can be derived for one $\beta$ only (they, of course, trivially extend to the elements in the orbit of $\beta$ under the irrational rotation by $\theta$) are not entirely satisfying since a quasicrystal is essentially modelled by the local isomorphism class of a given potential \cite{b}. It should therefore be expected that the dynamics behave uniformly with respect to $\beta$. Thus, our second goal here is to prove that $\alpha$-continuity indeed holds uniformly in $\beta$. Before stating the result, let us recall some basic notions from continued fraction expansion theory; we mention \cite{khin,lang} as general references.    
    
Given $\theta \in (0,1)$ irrational, we have an expansion    
$$    
\theta = \cfrac{1}{a_1+ \cfrac{1}{a_2+ \cfrac{1}{a_3 + \cdots}}}    
$$    
with uniquely determined $a_n \in \NN$. The associated rational    
approximants $\frac{p_n}{q_n}$ are defined by     
\begin{alignat*}{3}    
p_0 &= 0, &\quad        p_1 &= 1,   &\quad      p_n &= a_n p_{n-1} + p_{n-2},\\    
q_0 &= 1, &             q_1 &= a_1, &           q_n &= a_n q_{n-1} + q_{n-2}.    
\end{alignat*}    
The number $\theta$ is said to have bounded density if    
$$    
\limsup_{n \rightarrow \infty} \frac{1}{n} \sum_{i=1}^n a_i < \infty.    
$$    
The set of bounded density numbers is uncountable but has Lebesgue measure zero.    
    
\begin{theorem}\label{alphacont}    
Let $\theta$ be a bounded density number. Then for every $\lambda$ there exists $\alpha = \alpha (\lambda,\theta) > 0$ such that for every $\beta$, and every $\phi\in\ell^2(\ZZ)$ of compact support, the spectral measure for the pair $(H_{\lambda,\theta,\beta},\phi)$ is uniformly $\alpha$-H\"older continuous. In particular, $H_{\lambda,\theta,\beta}$ has purely $\alpha$-continuous spectrum.    
\end{theorem}    
    
\noindent{\it Remarks.}     
\begin{enumerate}    
    
\item A finite positive measure $d\Lambda$ is said to be uniformly $\alpha$-H\"older continuous (or U$\alpha$H) if the distribution    
function    
$$    
\Lambda(E)=\int_{-\infty}^Ed\Lambda    
$$    
is uniformly $\alpha$-H\"older continuous. A measure is said to be  
$\alpha$-continuous if it is absolutely continuous with respect to a U$\alpha$H measure.  This is    
equivalent to the statement $\mu(S)=0$ for all sets $S$ of zero $\alpha$-Hausdorff measure \cite{Ro}.    
    
\item See \cite{l} for explicit dynamical bounds that may be deduced from this result.    
    
\item As was noted in \cite{d1}, $\alpha$-continuity implies a lower bound on the Hausdorff dimension of the spectrum as a set. This lower bound has been complemented by Raymond in \cite{r}. This work contains a non-trivial upper bound on the dimension of the spectrum in the Fibonacci case at large coupling. We refer the reader to \cite{kkl} which proves upper bounds on the dynamics drawing on the ideas of \cite{r} and \cite{jl1}.    
\end{enumerate}    
    
The Gilbert-Pearson theory, as well as the Jitomirskaya-Last extension thereof, relates spectral measure properties to the asymptotic behavior of the following quantity,    
\begin{equation}\label{lnorm}    
\|u\|_L^2 = \sum_{n=0}^{\lfloor L \rfloor} \big|u(n)\big|^2 \; + \;(L-\lfloor L \rfloor)\big|u(\lfloor L \rfloor +1)\big|^2,    
\end{equation}    
where $u$ is a solution to (\ref{eve}). A proof of purely $\alpha$-continuous spectrum can be obtained by the following three-step procedure.    
    
\begin{itemize}    
\item Prove power-law upper bounds on $\|u\|_L$ for (spectrally almost) every energy $E$ in the spectrum of the given operator, uniformly for all solutions.    
\item Prove similar power-law lower bounds on $\|u\|_L$.    
\item Relate the spectral properties of the whole-line operator to the study of the solutions on the half-line.    
\end{itemize}    
The resulting $\alpha$ depends in an explicit way on the exponents in  
the power law bounds on $\|u\|_L$.   
   
Let us remark that some works, e.g. \cite{h,hks},    associate a slightly  
different family of operators to the parameters  $\lambda,\theta$  which is  
larger than the family parametrized by $\theta \in [0,1)$. The above  
theorems also hold for this family (cf. the corresponding discussions in \cite{dl1,dl2}).  
  
The organization of this article is as follows. In Section~2 we present some crucial properties of Sturmian potentials. We recall in particular the unique decomposition property and the uniform bounds on the traces of certain transfer matrices. Section~3 provides a study of the scaling properties of solutions to (\ref{eve}) with respect to the decomposition of the potentials on various levels and shows how Theorem \ref{cont} follows from these scaling properties. Uniform upper and lower power-law bounds on $\|u\|_L$ for certain rotation numbers are established in Section~4. Finally, Section~5 discusses the transition from half-line eigenfunction estimates to spectral properties of the whole-line operator which, together with the power-law bounds on the solutions, proves Theorem \ref{alphacont}. Since this interplay might be of independent interest, we present it in a general form, independent of the potential.    
    
\section{Basic properties of Sturmian potentials}    
    
In this section we recall some basic properties of Sturmian  
potentials. For further information we refer the reader to  
\cite{bist,d2,dl1,len,loth}.   We focus, in particular, on the decomposition of Sturmian potentials into canonical words, which obey recursive relations, and known results on the traces of the transfer matrices associated to these words.    
    
Fix some rotation number, $\theta$, and let $a_n$ denote the coefficients in its continued fraction expansion. Define the words $s_n$ over the alphabet $\mathcal{A}=\{0,1\}$ by    
\begin{equation}\label{recursive}    
s_{-1}^{} = 1, \quad s_0^{} = 0, \quad s_1^{} = s_0^{a_1 - 1} s_{-1}^{},    
\quad s_n^{} = s_{n-1}^{a_n} s_{n-2}^{}, \; n \ge 2.    
\end{equation}    
In particular, the word $s_n$ has length $q_n$ for each $n \ge 0$. By definition, $s_{n-1}$ is a prefix of $s_n$ for each    
$n \ge 2$. For later use, we recall the following elementary formula \cite{dl1}.    
    
\begin{prop}\label{wunderformel}    
For each $n \ge 2$, $s_n^{} s_{n+1}^{}= s_{n+1}^{} s_{n-1}^{a_n - 1} s_{n-2}^{} s_{n-1}^{}$.    
\end{prop}    
    
Thus, the word $s_n s_{n+1}$ has $s_{n+1}$ as a prefix. Note    
that the dependence of $a_n, p_n, q_n, s_n$ on $\theta$ is     
left implicit. Fix coupling constant $\lambda$ and energy $E$; then, for each $w=w_1\ldots w_n\in \mathcal{A}^n$, we    
define the transfer matrix $M(\lambda,E,w)$ by    
\begin{equation}\label{transfermatrices}    
M(\lambda,E,w) = \begin{bmatrix} E-\lambda w_n & -1\\1 & 0 \end{bmatrix} \times \cdots \times    
        \begin{bmatrix} E-\lambda w_1 & -1\\1 & 0 \end{bmatrix}.      
\end{equation}       
If $u$ is a solution to (\ref{eve}), we have    
$$    
U(n+1)=M\big(\lambda, E, v_{\theta,\beta}(1)\ldots v_{\theta,\beta}(n)\big)U(1),    
$$    
where    
$$    
U(n) = \begin{bmatrix} u(n)\\u(n-1) \end{bmatrix}.    
$$    
When studying the power-law behavior of $\|u\|_L$, one can investigate as well the behavior of    
\begin{equation}\label{LNorm}    
\|U\|_L = \left(\sum_{n=1}^{\lfloor L \rfloor} \big\|U(n)\big\|^2 \; +    
        \; (L-\lfloor L \rfloor)\big\|U(\lfloor L \rfloor +1)\big\|^2\right)^{\frac{1}{2}},    
\end{equation}    
where    
$$    
\|U(n)\|^2 = |u(n)|^2 + |u(n-1)|^2,    
$$    
since    
\begin{equation}\label{lleq}    
\tfrac{1}{2}\|U\|_L^2 \le \|u\|_L^2 \le \|U\|_L^2.    
\end{equation}

Now, the spectrum of $H_{\lambda,\theta,\beta}$ is independent of    
$\beta$ \cite{bist} and can thus be denoted by $\Sigma_{\lambda,\theta}$.     
Let us define    
\begin{align*}    
x_n &= \tr\big( M(\lambda,E,s_{n-1})\big), \\    
y_n &= \tr\big( M(\lambda,E,s_n)\big), \\    
z_n &= \tr\big(M(\lambda,E,s_n s_{n-1})\big),    
\end{align*}    
with dependence on $\lambda$ and $E$ suppressed.    
    
\begin{prop}\label{tracebound}    
For every $\lambda$ there exists $C_\lambda \in (1,\infty)$ such that for every irrational $\theta$, every $E \in \Sigma_{\lambda,\theta}$, and every $n \in \NN$, we have    
$$    
\max\, \{ |x_n|, |y_n|, |z_n| \} \leq C_\lambda.    
$$    
\end{prop}    
    
\noindent    
{\it Proof.} This result follows implicitly from \cite{bist}. It can be derived from the analysis in \cite{bist} by combining their bound on $|x_n|$ and $|y_n|$ with the fact that the traces obey the Fricke-$\!$Vogt invariant    
$$    
x_n^2 + y_n^2 + z_n^2 - x_n^{} y_n^{} z_n^{} = \lambda^2 + 4,    
$$    
which was also shown in \cite{bist}.\qed    
    
\medskip    
    
The words $s_n$ are now related to the sequences $v_{\theta,\beta}$ in the following way. For each pair    
$(\theta,n)$, every sequence $v_{\theta,\beta}$ may be partitioned into words such that each word is either    
$s_n$ and $s_{n-1}$.    
This uniform combinatorial property, together with the uniform trace bounds given in Proposition \ref{tracebound}, lies    
at the heart of the results contained in this paper and its precursors \cite{dl1,dl2}. Let us make this property explicit.    
\begin{definition}    
\rm    
Let $n\in \NN_0$ be given. An $(n,\theta)$-partition of a function $f:\ZZ \longrightarrow \{0,1\}$ is a sequence of pairs $(I_j, z_j)$, $j\in\ZZ$ such that:    
\begin{SmallList}    
\item the sets $I_j\subset \ZZ$ partition $\ZZ$;    
\item $1 \in I_0$;     
\item each block $z_j$ belongs to $\{s_n,s_{n-1}\}$; and    
\item the restriction of $f$ to $I_j$ is $z_j$. That is, $f_{d_j}f_{d_j +1}\ldots f_{d_{j+1}-1}=z_j$.    
\end{SmallList}                 
\end{definition}    
We will suppress the dependence on $\theta$ if it is understood to which $\theta$ we refer. In particular, we will write $n$-partition instead of $(n,\theta)$-partition. The unique decomposition property is now given in the following lemma which was proved in \cite{dl1}.    
    
\begin{lemma}\label{partition-lemma}    
For every $n\in \NN_0$ and every $\beta \in [0,1)$, there exists a unique $n$-partition $(I_j,z_j)$ of $v_{\theta,\beta}$. Moreover, if $z_j=s_{n-1}$, then $z_{j-1}=z_{j+1}=s_n$. If $z_j=s_n$, then there is an interval $I=\{d,d+1,\ldots,d+l-1\}\subset \ZZ$    
containing $j$ and of length $l\in\{a_{n+1},a_{n+1}+1\}$ such that $z_i=s_n$ for all $i\in I$ and    
$z_{d-1}=z_{d+l}=s_{n-1}$.    
\end{lemma}    
  
We finish this section with a short discussion of symmetry properties of   
the words $v_{\theta,\beta}$. This will show that the considerations  
below, based on a study of the operators $H_{\lambda,\theta,\beta}$ on the right half-line, could equally well be based on a study of the operators on the left half-line. This particularly implies that for all parameter values, given an energy in the spectrum, both at $+\infty$ and $- \infty$ every solution of (\ref{eve}) does not tend to zero.  
  
For a finite word $w=w_1\ldots w_n$ over $\{0,1\}$, define the reverse word  
$w^R$ by $w^R = w_n\ldots w_1$ and for a word $w\in \{0,1\}^\ZZ$, define the reverse word $w^R$ by $w^R=v$ with $v_n=w_{-n}$ for $n\in \ZZ$. It is not hard to show that every $v_{\theta,\beta}$ allows a unique $n$-$R$-partition \cite{len}. Here, an $n$-$R$-partition is defined by replacing $s_{n-1}$ and $s_n$ by $s_{n-1}^R$ and $s_n^R$, respectively, in the definition of $n$-partition. Mimicking the proof of Lemma 5.1 in \cite{dl2} with the norm replaced by the trace, gives immediately $x_n^R=x_n$, $y_n^R=y_n$ and $z_n^R=z_n$. Here, $x_n^R,y_n^R$ and $z_n^R$ are defined by replacing $s_{n-1}, s_n$ and $s_n s_{n-1}$ with their reverse words in the definition of $x_n,y_n$ and $z_n$, respectively.  Thus, the analog of Proposition \ref{tracebound} holds for $x_n^R, y_n^R, z_n^R$ (in fact, this can also be established by remarking that the underlying trace map system is  
essentially unchanged by passing from $s_n$ to $s_n^R$).  The  
$n$-$R$-partitions and the bound on the traces allow to study the operators  
on the left half-line in exactly the same way as the operators on the  
right half-line are studied in the following two sections. Alternatively, it is possible to show that the map $R$ leaves the set $\overline{\{v_{\theta,\beta} \,:\, \beta \in [0,1)\}} \subset \{0,1\}^\ZZ$ invariant, where the bar denotes closure with respect to product topology \cite{len}. This could also be used to show that the two half-lines are equally well accessible.  
    
\section{Scaling behavior of solutions}    
    
In this section, we use the trace bounds and the partition lemma to study the growth of $\|U\|_{L}$ for energies in the spectrum and normalized solutions to (\ref{eve}). For our purposes it will be sufficient    
to consider this quantity only for $L=q_{8n}$, $n \in \NN$.  In Lemma \ref{scaling} below it is shown that this growth has a lower bound which is exponential in $n$. In particular, this will imply absence of eigenvalues as claimed in Theorem \ref{cont} and it will also be used in our proof of power-law (in $L$) lower bounds for certain rotation numbers which will be given in the next section.    
    
\begin{lemma}\label{scaling}    
Let $\lambda, \theta, \beta$ be arbitrary, $E \in  
\Sigma_{\lambda,\theta}$, and let $u$ be a normalized solution to  
\eqref{eve}. Then, for every $n \ge 8$, the inequality    
$$    
\|U\|_{q_n} \ge D_\lambda \|U\|_{q_{n-8}},    
$$    
holds, where    
$$    
D_\lambda^2 = 1 + \big[\tfrac{1}{2 C_\lambda}\big]^2.    
$$    
\end{lemma}    
    
\medskip    
    
\noindent{\it Proof of Theorem \ref{cont}.} It follows immediately from Lemma \ref{scaling} that for all parameter values $\lambda, \theta, \beta$, the operator $H_{\lambda, \theta, \beta}$ has no eigenvalues.\qed     
    
\medskip    
    
Before giving the proof of Lemma \ref{scaling}, let us make explicit the core argument we shall use. Similar to \cite{d1}, we employ a mass-reproduction technique which is based upon the two-block version of the Gordon argument from \cite{g2}.    
    
\begin{lemma}\label{gordoncrit}    
Fix $\lambda, \theta, \beta$. Suppose that $v_{\theta,\beta}(j) \ldots  
v_{\theta,\beta}(j+2k-1)$ is conjugate to $(s_{n-1})^2$, $(s_n)^2$, or  
$(s_{n-1}s_n)^2$ for some $n \in \NN$, $l \le k$, and every $j \in  
\{1,\ldots,l\}$. Let $E \in \Sigma_{\lambda, \theta}$. Then every  
normalized solution $u$ to \eqref{eve} satisfies  
$$    
\|U\|_{l+2k} \ge D_\lambda \|U\|_l.    
$$    
\end{lemma}    
\noindent{\it Remark.} A word $w = w_1 \ldots w_n$ is conjugate to a word $v = v_1 \ldots v_n$ if for some $i \in \{ 1, \ldots , n\}$, we have $w_1 \ldots w_n = v_i \ldots v_n v_1 \ldots v_{i-1}$, that is, if $w$ is obtained from $v$ by a cyclic permutation of its symbols.    
    
\medskip    
    
\noindent {\it Proof.} Consider some $j \in \{1,\ldots,l\}$. By definition, we have    
$$    
U(j+k) = M\big(\lambda,E,v_{\theta,\beta}(j) \ldots v_{\theta,\beta}(j+k-1)\big) U(j),    
$$    
and by assumption,    
\begin{align*}    
U(j+2k) &= M\big(\lambda,E,v_{\theta,\beta}(j) \ldots v_{\theta,\beta}(j+2k-1)\big) U(j)\\    
&= \left[ M\big(\lambda,E,v_{\theta,\beta}(j) \ldots v_{\theta,\beta}(j+k-1)\big) \right]^2 U(j).    
\end{align*}    
Hence, applying the Cayley-Hamilton Theorem,    
\begin{equation}\label{reprep}    
U(j+2k) - \tr \big[M\big(\lambda,E,v_{\theta,\beta}(j) \ldots v_{\theta,\beta}(j+k-1)\big)\big] U(j+k) + U(j) = 0.    
\end{equation}    
Moreover,    
\begin{equation}\label{tb}    
\big| \tr \big[M\big(\lambda,E,v_{\theta,\beta}(j) \ldots v_{\theta,\beta}(j+k-1)\big)\big] \big| \leq C_\lambda.    
\end{equation}    
Combining (\ref{reprep}) and (\ref{tb}), we obtain     
\begin{equation}\label{lb}    
\max\:\big\{ \|U(j+k)\| , \|U(j+2k)\| \big\} \geq \frac{1}{2C_\lambda} \|U(j)\|    
\end{equation}    
for all $1 \leq j \leq l$. We can therefore proceed as follows,    
\begin{align*}    
\|U\|_{l+2k}^2 &= \sum_{m=1}^{l+2k} \|U(m)\|^2\\    
&=   \sum_{m=1}^{l} \|U(m)\|^2\; + \sum_{m=l+1}^{l+2k} \|U(m)\|^2\\    
&\ge \sum_{m=1}^{l} \|U(m)\|^2 + \big[ \tfrac{1}{2C_\lambda} \big]^2 \sum_{m=1}^{l} \|U(m)\|^2\\    
&=   \Big( 1 + \big[ \tfrac{1}{2 C_\lambda}\big]^2 \Big) \|U\|_l^2.    
\end{align*}    
This proves the assertion.\qed    
    
\medskip    
    
\noindent{\it Proof of Lemma \ref{scaling}.} We make use of the information provided by Lemma \ref{partition-lemma} and exhibit squares in the potentials which are suitable in the sense that they satisfy the assumption of Lemma \ref{gordoncrit}. In fact, we shall show    
\begin{equation}\label{goal}    
\|U\|_{2(q_{n+1} + q_n) + q_{n-1}} \ge D_\lambda \|U\|_{q_{n-4}}    
\end{equation}    
for all $\lambda,\theta,\beta$, all $E \in \Sigma_{\lambda,\theta}$, all solutions $u$, and all $n \ge 4$. Since $q_{n+4} \ge 2(q_{n+1} + q_n) + q_{n-1}$, this proves the assertion.    
    
Fix $\lambda, \theta, \beta$ and some $n \ge 4$ and consider the $n$-partition of $v_{\theta,\beta}$. Since we want to exhibit squares close to the origin, we consider the following cases.    
\begin{CaseHier}{3mm}    
    
\item[\it Case 1: $z_0 = s_{n-1}$. ] Applying (\ref{recursive}) and Proposition \ref{wunderformel}, we see that this block is followed by $s_{n-1}^2 s_{n-4}^{}$. We can, therefore, apply Lemma \ref{gordoncrit} with $l = q_{n-4}$ and $k = q_{n-1}$. This yields (\ref{goal}) and we are done in this case.    
    
\item[\it Case 2: $z_0  = s_n$ and $z_1 = s_n$. ] Proposition (\ref{wunderformel}) yields that these two blocks are followed by $s_n s_{n-3}$. Lemma \ref{gordoncrit} now applies with $l = q_{n-3}$ and $k = q_n$.    
    
\item[\it Case 3: $z_0  = s_n$ and $z_1 = s_{n-1}$. ] Let $z_j'$ label the blocks in the $(n+1)$-partition of $v_{\theta, \beta}$. Therefore we have $z_0' = s_{n+1}$. Let us consider the following subcases.    
\begin{CaseHier}{1.5mm}    
    
\item[\it Case 3.1: $z_1' = s_{n+1}$. ] Similar to Case 2, this implies that $z_0' z_1'$ is followed by $s_{n+1}s_{n-2}$ and hence Lemma \ref{gordoncrit} applies with $l = q_{n-2}$ and $k = q_{n+1}$.    
    
\item[\it Case 3.2: $z_1' = s_n$. ] It follows that $z_2' = s_{n+1}$. Again we consider two subcases.    
\begin{CaseHier}{0mm}    
    
\item[\it Case 3.2.1: $z_3' = s_n$. ] Of course, this case can only occur if $a_{n+2} = 1$. We infer that $z_4' = s_{n+1}$. But this implies that we have squares conjugate to $s_n s_{n+1}$ and Lemma~\ref{gordoncrit} is applicable with $l = q_{n-1}$ and $k = q_n + q_{n+1}$. Hence, (\ref{goal}) also holds in this case.    
    
\item[\it Case 3.2.2: $z_3' = s_{n+1}$. ] Let us consider the consequences of this particular case for the blocks in the $n$-partition. We have    
\begin{equation}\label{case322}    
z_0 z_1 \ldots z_{2 a_{n+1} + 4} = s^{}_n s^{}_{n-1} s^{}_n s^{a_{n+1}}_n s^{}_{n-1} s^{a_{n+1}}_n s^{}_{n-1}.    
\end{equation}     
Since $s_n$ is a prefix of $s_{n+1}$, this must be followed by $s_n$. We therefore have the sequence of blocks    
$$    
s_n^{} s_{n-1}^{} s_n^{} s_n^{a_{n+1}} s_{n-1}^{} s_n^{a_{n+1}} s_{n-1}^{} s_n^{}    
$$    
where the site $1 \in \ZZ$ is contained in the leftmost block. Using Proposition \ref{wunderformel} this can be rewritten as    
$$    
s_n^{} s_{n-1}^{} s_n^{} s_n^{a_{n+1}} s_{n-1}^{} s_n^{a_{n+1}} s_n^{} s_{n-2}^{a_{n-1}-1} s_{n-3}^{} s_{n-2}^{},    
$$    
which can as well be interpreted as    
$$    
s_n^{} s_{n-1}^{} s_n^{} s_n^{a_{n+1}} s_{n-1}^{} s_n^{} s_n^{a_{n+1}} s_{n-2}^{a_{n-1}-1} s_{n-3}^{} s_{n-2}^{}.    
$$    
Thus, Lemma \ref{gordoncrit} is applicable with $l = q_{n-3}$ and $k = q_n + q_{n+1}$ which closes Case~3.2.2.    
\end{CaseHier}    
\end{CaseHier}    
\end{CaseHier}    
\vspace{-1.5mm}    
Between Cases 1, 2, and 3 we have covered all possible choices of $z_0,z_1$.\qed    
    
\section{Power-law upper and lower bounds on solutions}    
    
In this section we provide power-law bounds for $\|u\|_L$ in the case where the rotation number $\theta$ has suitable number theoretic properties. Recall that $a_n$ denote the coefficients in the continued fraction expansion of $\theta$ and $q_n$ denote the denominators of the canonical continued fraction approximants to $\theta$.

\begin{prop}\label{lowerpower}    
Let $\theta$ be such that for some $B < \infty$, $q_n \le B^n$ for every $n \in \NN$. Then for every $\lambda$, there exist $0 < \gamma_1, C_1 < \infty$ such that for every $E \in \Sigma_{\lambda,\theta}$ and every $\beta$, every solution $u$ of \eqref{eve},\eqref{norm} obeys    
\begin{equation}\label{lpb}    
\|u\|_L \ge C_1 L^{\gamma_1}    
\end{equation}    
for $L$ sufficiently large.    
\end{prop}    
\noindent{\it Remark.} The set of $\theta$'s obeying the assumption of Proposition \ref{lowerpower} has full Lebesgue measure \cite{khin}.    
    
\medskip    
    
\noindent{\it Proof.} The bound (\ref{lpb}) can be derived from the exponential lower bound on $\|U\|_{q_{8n}},{n \in \NN}$ given the exponential upper bound on $q_n,{n \in \NN}$. Lemma \ref{scaling} established the power-law bound for $L=q_{8n}$. It can then be interpolated to other values of $L$ (see \cite{d1} for details).\qed

\begin{prop}\label{upperpower}    
Let $\theta$ be a bounded density number. Then for every $\lambda$, there exist $0 < \gamma_2, C_2 < \infty$ such that for every $E \in \Sigma_{\lambda,\theta}$ and every $\beta$, every normalized solution $u$ of \eqref{eve},\eqref{norm} obeys    
\begin{equation}\label{upb}    
\|u\|_L \le C_2 L^{\gamma_2}    
\end{equation}    
for all $L$.    
\end{prop}    
\noindent{\it Proof.} The proof is based upon local partitions and results by Iochum et al.~\cite{irt,it}. Up to interpolation to non-integer $L$'s, it was given in \cite{dl2}.\qed\\[5mm]    
{\it Remark.} It is easy to see that bounded density numbers obey the assumption of Proposition \ref{lowerpower}. Thus, if $\theta$ is a bounded density number, we have     
$$    
C_1 L^{\gamma_1} \le \|u\|_L \le C_2 L^{\gamma_2}    
$$    
with $\lambda$-dependent constants $\gamma_i,C_i$, uniformly for all energies from the spectrum, all phases $\beta$, and all normalized solutions of \eqref{eve}.    
    
\section{Subordinacy Theory}    
    
In this section we demonstrate how the solution estimates of the previous section may be used to prove $\alpha$-continuity of spectral measures for some $\alpha>0$.    
    
As it will cost us nothing in clarity, we shall treat the operator    
\begin{equation*}    
[Hu](n) =  u(n+1) + u(n-1) + V(n)u(n)     
\end{equation*}    
with arbitrary potential $V\!\!:\!\ZZ\to\RR$. To each such whole-line operator we associate two half-line operators, $H_+=P_+^* H P_+$ and $H_-=P_-^* H P_-$, where $P_\pm$ denote the inclusions    
$P_+\!:\!\ell^2(\{1,2,...\})\hookrightarrow\ell^2(\ZZ)$ and     
$P_-\!:\!\ell^2(\{0,-1,-2,...\})\hookrightarrow\ell^2(\ZZ)$.    
    
The spectral properties of $H,H_\pm$ are typically studied via the Weyl $m$-functions.    
For each $z\in\CC\setminus\RR$ we define $\psi^\pm(n;z)$ to be the unique solutions to    
$$    
H\psi^\pm=z\psi^\pm,\quad \psi^\pm(0;z)=1\quad\text{and}\quad\sum_{n=0}^\infty |\psi^\pm(\pm n;z) |^2 < \infty.    
$$    
With this notation we can define the Weyl functions by    
\begin{align*}    
m^+(z) &= \langle\delta_1 | (H_+-z)^{-1} \delta_1 \rangle = -\psi^+(1;z)/\psi^+(0;z)\\    
m^-(z) &= \langle\delta_0 | (H_--z)^{-1} \delta_0 \rangle = -\psi^-(0;z)/\psi^-(1;z)    
\end{align*}    
for each $z\in\CC\setminus\RR$.    
Here and elsewhere, $\delta_n$ denotes the vector in $\ell^2$ supported at $n$ with $\delta_n(n)=1$.    
For the whole-line problem the $m$-function role is played by the $2\times2$ matrix $M(z)$:    
$$    
\left[\begin{smallmatrix}  a \\  b \end{smallmatrix}\right]^\dagger M(z)    
        \left[\begin{smallmatrix}  a \\  b \end{smallmatrix}\right]    
= \big\langle (a\delta_0+b\delta_1)\big|(H-z)^{-1} (a\delta_0+b\delta_1) \big\rangle.    
$$    
Or, more explicitly,    
\begin{align*}    
M &= \frac{1}{\psi^+(1)\psi^-(0)-\psi^+(0)\psi^-(1)}    
        \begin{bmatrix} \psi^+(0)\psi^-(0) & \psi^+(1)\psi^-(0) \\    
                \psi^+(1)\psi^-(0) & \psi^+(1)\psi^-(1) \end{bmatrix} \\    
  &= \frac{1}{1-m^+m^-}    
                \begin{bmatrix} m^- & -m^+m^- \\    
                -m^+m^- & m^+ \end{bmatrix}    
\end{align*}    
with $z$ dependence suppressed. We define $m(z)=\tr\big(M(z)\big)$, that is, the trace of $M$. From these definitions one obtains:    
\begin{align}    
\notag    
m^{\pm}(z) &= \int \frac{1}{t-z} d\rho^{\pm}(t), \\    
\label{4:mRep}    
m(z) &= \int \frac{1}{t-z} d\Lambda(t),    
\end{align}    
where $d\rho^{+},d\rho^{-}$ are the spectral measures for the pairs $(H_+,\delta_1),(H_-,\delta_0)$, respectively, and $d\Lambda$ is the sum of the spectral measures for the pairs $(H,\delta_0)$ and $(H,\delta_1)$. An immediate consequence of these representations is that each of the $m$-functions maps $\CC^+=\{x+iy:y>0\}$ to itself.    
    
The pair of vectors $\{\delta_0,\delta_1\}$ is cyclic for $H$; indeed, if $\phi$ is supported in $\{-N,\ldots,N,N+1\}$, then there exist polynomials $P_0,P_1$ of degree not exceeding $N$ such that $\phi=P_0(H)\delta_0+P_1(H)\delta_1$. This may be proved readily, by induction, once it is observed that $\phi(-N),\phi(N+1)$ uniquely determine the leading coefficients of $P_0,P_1$, respectively.    
    
Our immediate goal is to prove that $d\Lambda$ is uniformly $\alpha$-H\"older continuous. This will follow quickly from    
    
\begin{theorem}\label{th3}    
Fix $E\in\RR$. Suppose every solution of $(H-E)u=0$ with $|u(0)|^2+|u(1)|^2=1$ obeys the estimate    
\begin{equation}    
\label{4:solns}    
C_1L^{\gamma_1} \leq \|u\|_L \leq C_2L^{\gamma_2}    
\end{equation}    
for $L>0$ sufficiently large. Then    
\begin{equation}    
\label{4:mEst}    
\sup_\varphi \left| \frac{\sin(\varphi)+\cos(\varphi)m^+(E+i\epsilon)}    
                {\cos(\varphi)-\sin(\varphi)m^+(E+i\epsilon)} \right|    
        \leq C_3 \epsilon^{\alpha-1},    
\end{equation}    
where $\alpha=2\gamma_1/(\gamma_1+\gamma_2)$.    
\end{theorem}    
\noindent{\it Proof.} This result lies within the Gilbert-Pearson theory of subordinacy \cite{gp,g1,kp}. A concise proof is available in \cite{jl1,jl2}. In this context, the $\varphi$ above corresponds to the choice of boundary    
conditions.\qed    
    
\begin{coro}\label{piac}    
Given a Borel set $\Sigma$, suppose that the estimate \eqref{4:solns} holds for every $E\in\Sigma$ with $C_1,C_2$ independent of $E$.    
Then, given any function $m^-\!:\!\CC^+\to\CC^+$, and any $E\in\Sigma$,    
\begin{equation}    
\label{m:n}    
|m(E+i\epsilon)| =    
        \left| \frac{m^+(E+i\epsilon)+m^-(E+i\epsilon)}{1-m^+(E+i\epsilon)m^-(E+i\epsilon)} \right|    
        \leq C_3\,\epsilon^{\alpha-1}    
\end{equation}    
for all $\epsilon>0$. Consequently, $\Lambda(E)$ is uniformly  
$\alpha$-H\"older continuous at all points $E\in\Sigma$. In particular, $d\Lambda$ is $\alpha$-continuous on $\Sigma$.    
\end{coro}    
    
\noindent    
{\it Proof}. Fix $E\in\Sigma$ and $\epsilon>0$. Then, by introducing new variables $z=e^{2i\varphi}$ and $\mu=(m^+-i)/(m^++i)$, we may rewrite (\ref{4:mEst}) as    
$$    
\sup_{|z|=1} \left| \frac{1+\mu z}{1-\mu z} \right| \leq C_3 \epsilon^{\alpha-1}.    
$$    
Note that $\Im(m^+)>0$ implies $|\mu|<1$ and so $(1+\mu z)/(1-\mu z)$ defines an analytic function on $\{z:|z|\leq 1\}$. The point $z=(i-m^-)/(i+m^-)$ lies inside the unit disk since $\Im(m^-)>0$. The estimate (\ref{m:n}) now follows from the maximum modulus principle and a few simple manipulations. This estimate and the representation \eqref{4:mRep} provide    
$$    
\Lambda\big([E-\epsilon,E+\epsilon]\big) \leq 2\epsilon\Im\big(m(E+i\epsilon)\big)\leq 2C_3\,\epsilon^\alpha    
\quad \text{for all $E\in\Sigma$, $\epsilon>0$,}    
$$    
from which $\Lambda(E)$ is uniformly $\alpha$-H\"older continuous on $\Sigma$.\qed    
    
\medskip    
    
\noindent    
{\it Remark.} If we permit $C_1,C_2$ to depend on $E$, the only consequence is that now $C_3$ depends on $E$ and so $\Lambda$ need not be {\it uniformly}\ H\"older continuous. However, $\alpha$-continuity is     
still guaranteed.    
    
\medskip    
    
\noindent    
{\it Proof of Theorem \ref{alphacont}.} Propositions \ref{lowerpower} and \ref{upperpower} provide the estimate (\ref{4:mEst}) for each $E$ in the spectrum $\Sigma_{\lambda,\theta}$ of $H_{\lambda,\theta,\beta}$.    
Of course $d\Lambda$ is supported by $\Sigma_{\lambda,\theta}$ and so must be uniformly $\alpha$-H\"older continuous.    
    
Given $\phi\in\ell^2(\ZZ)$ with compact support, the remarks preceding Theorem \ref{th3} show that the spectral measure for $\phi$ is bounded by $f(E)d\Lambda(E)$ with $f(E)$ uniformly bounded on the compact set $\Sigma$, which completes the proof.\qed    
    
\medskip    
    
{\it Acknowledgments.} D.~D.~was supported by the German Academic    
Exchange Service through Hochschulsonderprogramm III (Postdoktoranden)    
and D.~L.~received financial support from Studienstiftung des Deutschen    
Volkes (Doktorandenstipendium), both of which are gratefully acknowledged.

\end{document}